\documentclass[12pt,hyper]{article}
\usepackage{jhep-pub}
\usepackage{bm}
\usepackage{amssymb,amsmath,amsthm}
\usepackage{mathrsfs} 
\def\TODAY{11 May 2011; 15 August 2011}
\title{\bf From dispersion relations to spectral dimension --- and back again}
\author{Thomas P. Sotiriou$^{1,2}$, Matt Visser$^{3}$,  {\rm and} Silke Weinfurtner$^1$}

\affiliation{$^1$ SISSA - International School for Advanced Studies\\
via Bonomea 265,
34136 Trieste, Italy,
\emph{and}\\ INFN, Sezione di Trieste.}
\affiliation{$^2$ 
Department of Applied Mathematics and Theoretical Physics,
CMS\\
University of Cambridge, Wilberforce Road, Cambridge CB3 0WA, UK} 
\affiliation{$^3$ School of Mathematics, Statistics, and Operations Research\\
Victoria University of Wellington, PO Box 600, Wellington 6140, New Zealand}

\emailAdd{sotiriou@sissa.it}
\emailAdd{matt.visser@msor.vuw.ac.nz}
\emailAdd{silke.weinfurtner@sissa.it}

\abstract{
The so-called spectral dimension is a scale-dependent number associated with both geometries and field theories that has recently attracted much attention, driven largely, though not exclusively, by investigations of causal dynamical triangulations  (CDT) and Ho\v{r}ava gravity as possible candidates for quantum gravity. We advocate the use of the spectral dimension as a probe for the \emph{kinematics} of these (and other) systems in the region where spacetime curvature is small, and the manifold is flat to a good approximation. In particular, we show how to assign a spectral dimension (as a function of so-called diffusion time) to any arbitrarily specified dispersion relation. We also analyze the fundamental properties of spectral dimension using extensions of the usual Seeley--DeWitt and Feynman expansions, and by saddle point techniques.  The spectral dimension turns out to be a useful, robust and powerful probe, not only of geometry, but also of kinematics.

\bigskip
\noindent
Published as: Physical Review D {\bf84} (2011) 104018

\bigskip
\noindent
doi: 10.1103/PhysRevD.84.104018

\bigskip
\noindent
\TODAY; Published 8 November 2011; \\
\LaTeX-ed \today.
}

\enlargethispage{40pt}
\begin{document}
\maketitle
\flushbottom

\def\lint{\hbox{\Large $\displaystyle\int$}} 
\def\hint{\hbox{\Huge $\displaystyle\int$}}  
\def\d{{\mathrm{d}}}
\newcommand{\scri}{\mathscr{I}}
\newcommand{\sun}{\ensuremath{\odot}}
\def\d{{\mathrm{d}}}
\def\J{{\mathscr{J}}}
\def\L{{\mathscr{L}}}
\def\H{{\mathscr{H}}}
\def\T{{\mathscr{T}}}
\def\V{{\mathscr{V}}}
\def\sech{{\mathrm{sech}}}
\def\tr{{\mathrm{tr}}}
%

\section{Introduction}
\def\Horava{Ho\v{r}ava}

 The spectral dimension has been proposed as possible observable characterizing the geometry in discrete quantum gravity approaches \cite{CDT5}. It can be viewed as an effective notion of dimension defined though a fictitious diffusion process on the discrete geometry. In simple terms (see reference~\cite{CDT5} for more details),  the diffusion process can be thought as a stochastic random walk and the spectral dimension is defined in terms of the average return propability $P(s)$ of the fictitious diffusion as 
\begin{equation}
d_S(s) = -2 {\d \ln P(s)\over\d\ln s},
\end{equation}
with $s$ being the (fictitious) diffusion time. As opposed to the topological dimension $d$,  the spectral dimension $d_S(s)$ need not be an integer and furthermore is scale-dependent.

The concept of spectral dimension has attracted considerable interest in causal dynamical triangulations (CDT) \cite{CDT5, CDT1, CDT2, CDT3, CDT4, CDT6}, since finding meaningful observables in discrete theories is not a trivial task. 
(See also references~\cite{Benedetti:2008gu,Laiho:2011ya,Atkin:2011ak} where the spectral dimension is used in noncommutative spacetimes, Euclidean dynamical triangulations and random combs respectively.)
Such observables can provide the much needed connection with continuum theories (at suitable limits). Indeed the spectral dimension can be defined in the continuum quantum gravity models as well and be used to characterize and understand their short-distance behaviour (see references~\cite{Horava, Carlip1, Carlip2, Modesto:2009qc, Pinzul:2010ct}). Furthermore, it was shown in reference~\cite{Horava} that both CDT and Ho\v rava gravity seem to lead to a spectral dimension of 2 in the ultraviolet limit, while the value of the spectral dimension matches that of the topological dimension at larger length scales. These first results suggest that the spectral dimension can be a useful potential link between discrete and continuum theories.

Clearly if one is to utilize the spectral dimension in this way, it is crucial to first get a deep understanding one what kind of information it actually carries in continuum theories and also develop a set of tools for extracting this information. Our goal here is to make a first step in this direction.

Given the definition of a spectral dimension, it is actually quite intuitive (and will become clearer shortly) that, in continuum theories, its behaviour is scale-dependent and one can expect that:
\begin{enumerate}

\item 
At long distances, scales  which are comparable to, or larger than, the radius of curvature, the spectral dimension will depend on the details of the geometry and the presence gravitational sources. At these scales the deviation of the spectral dimension from the topological dimension can be essentially attributed to curvature effects. Alternatively, at these scales the spectral dimension probes aspects of the geometry.

\item At intermediate scales, large compared to the Planck scale but small compared to the radius of curvature, spacetime is effectively flat, and one should obtain a spectral dimension whose value matches that of the usual topological dimension of spacetime.

\item At short distances, (typically Planckian distances, but sometimes one is instead interested in the near-horizon limit of a black hole), curvature effects can certainly be neglected, and the  deviation of the spectral dimension from the topological dimension should be attributed to different effects. 

\end{enumerate}
We will not be directly interested in regime 1 for the purposes of the current analysis. That is, we will not attempt any thorough analysis of how the spectral dimension can be used as a probe of geometry or how curvature affects the spectral dimension (though we will comment upon it briefly). Instead, in this article we shall focus on the flow of the spectral dimension from flat spacetime, which is for our purposes the infrared limit,  into the ultraviolet limit (regime 2 above to regime 3 above). The corresponding length scales are of particular interest in quantum gravity candidates.  We shall develop a number of theoretical results, and complement them with analytic and computational tools that allow for both exact and approximate evaluation of the spectral dimension in a number of interesting situations. 

In particular, we shall demonstrate that the spectral dimension is \emph{not necessarily} intrinsically geometric. At scales small enough for curvature effects to be negligible its deviations from the topological dimension it actually becomes an \emph{analytic} property of the differential operator (ultimately equivalent to a dispersion relation) that one is using as input to define the generalized diffusion process (as already suggested by the analysis of reference~\cite{Horava} for the particular case of Ho\v rava gravity). In turn, this operator acts as the propagator of some dynamical degree of freedom in flat space. In this sense, the spectral dimension acts, at suitable scales, as a probe of the kinematics of the particular degree of freedom. Within this framework we will show that, given a specified topological dimension $d$, it is possible to define a scale-dependent notion of a spectral dimension $d_S(s)$ for \emph{any} arbitrary dispersion relation $\omega = \Omega(k)$. We shall also derive a formal  inversion process, so that from the spectral dimension, (as a function of $s$),  and the topological dimension $d$, one can in principle reconstruct the dispersion relation. 

For some specific dispersion relations the situation is sufficiently simple that it is possible to explicitly evaluate the relevant integrals, and so explicitly determine the spectral dimension in terms of standard special functions (such as Bessel functions and [Gaussian] error functions). 
More generally,  we shall develop a number of analytic tools, (using a generalization of the Seeley--DeWitt expansion in the ultraviolet, and a generalization of the Feynman expansion in the infrared), that give us good analytic control of these two limits.  We also develop an approximation based on saddle point techniques to express the spectral dimension in terms of the phase and group velocities of the underlying dispersion relation --- this approximation agrees with asymptotic results in the ultraviolet and infrared, and can also be used in the intermediate regime. (When saddle point and exact integral techniques overlap, the saddle point approximation will be seen to be a tolerably good global fit to the exact result.) 

Apart from those issues of relevance to the quantum gravity and field theory communities, we additionally point out that the formalism derived in this article may also be of relevance in completely non-relativistic situations (such as the propagation of surface waves on liquid-gas interfaces.)  Overall, we wish to advocate the use of spectral dimension as a general, powerful, and robust probe of the kinematics (specifically the dispersion relation) of whatever system one has under consideration.

\section{Theoretical  framework}
\subsection{Dispersion relation}

Suppose we are working in flat $(d+1)$ dimensional spacetime and have a (in general Lorentz-violating) dispersion relation
\begin{equation}
\omega = \Omega(k).
\end{equation}
Now define
\begin{equation}
f(k^2) = \Omega(k)^2.
\end{equation}
This dispersion relation can always be viewed as being completely specified by the solutions of the (in general Lorentz-violating) differential equation
\begin{equation}
D_L \Phi \equiv  \left\{ - \partial_t^2 - f(-\nabla^2) \right\} \Phi = 0\,.
\end{equation}
The reason that we choose the differential operator to be second order in physical time $t$ is a purely pragmatic one --- with such a choice the differential equation encoding the dispersion relation can typically be derived from a \emph{ghost free} non-interacting  Lagrangian
\begin{equation}
\L = {1\over2} \; \Phi D_L \Phi = {1\over2} \; \Phi  \left\{ - \partial_t^2 - f(-\nabla^2) \right\} \Phi .
\end{equation}

\subsection{Spectral dimension}

First we Wick-rotate in physical time $t$ to consider the Euclideanized differential operator $D_E$ in $(d+1)$ topological dimensions
\begin{equation}
D_E \Phi \equiv  \left\{ - \partial_t^2 + f(-\nabla^2) \right\} \Phi.
\end{equation}
Now consider the generalized diffusion process 
\begin{equation}
\label{diffusion}
\left\{ {\partial\over\partial s} + D_E \right\} \rho(x,x', s) = 0; \qquad   \rho(x,x',0) = \delta^{d+1}(x-x'), 
\end{equation}
where $D_E$ is the differential operator defined above. We set $x = (t,\vec x)$, and $s$ is best thought of as an auxiliary diffusion time. (That is, $s$ is a ``fake time'',  a Lorentz-violating generalization of Schwinger--De~Witt proper time --- as in the usual Schwinger--De~Witt proper-time regularization of spacetime propagators.) We shall soon see that $s$ can also be thought of as the scale at which one is probing the diffusion process --- we shall see that $s\to0$ probes the ultraviolet, while $s\to \infty$ probes the infrared.
The general solution of equation~(\ref{diffusion}) is given by
\begin{equation}
\rho(x,x',s) = \int {\d^d k \; \d \omega \over(2\pi)^{d+1}}   \exp\{ i [\vec k \cdot (\vec x-\vec x') +\omega(t-t')] \} \exp\{-s[\omega^2+ f(k^2) ]\},
\end{equation}
as can easily be verified by differentiation and by noting that the boundary condition is satisfied at $s=0$.  The quantity $ \rho(x,x', s)$ is often called the heat kernel and can be thought of as the probability density that a particle originally at $x'$ at $s=0$ will diffuse to $x$ in diffusion time $s$. 
Now consider the  average return probability density $P(s)$  (also called the diagonal part of the heat kernel):
\begin{equation}
P(s) = \rho(x,x,s) =  \int {\d^d k \; \d \omega \over(2\pi)^{d+1}} \exp\{-s[\omega^2+ f(k^2) ]\},
\end{equation}
(We are working in flat space where, due to translational invariance the average return probability equals the return probability.)
The spectral dimension is defined as
\begin{equation}
\label{dsdef}
d_S(s) = -2 {\d \ln P(s)\over\d\ln s}=  -2 {\d \ln \rho(x,x;s)\over\d\ln s}.
\end{equation}
The reason for this definition is that if we consider the particularly simple differential operator $D= \Box = -\nabla^2$, then in $(d+1)$ topological dimensions then one can show that
\begin{equation}
P_{-\nabla^2}(s) = \rho_{-\nabla^2}(x,x;s) \propto s^{-(d+1)/2}
\end{equation}
so that in this specific case
\begin{equation}
d_{S;-\nabla^2}(s) = -2 {\d \ln P_{-\nabla^2}(s)\over\d \ln s}=  -2 {\d \ln \rho_{-\nabla^2}(x,x;s)\over\d \ln s} = d+1.
\end{equation}
That is: The spectral dimension $d_{S;D_E}(s)$ of the complicated diffusion process governed by the differential operator $D_E$ is the (fictitious) topological dimension of that simple diffusion process governed by $-\nabla^2$ that most closely approximates $P_{D_E}(s)$ at the indicated value of the diffusion time $s$. We emphasise that this is a definition, not a theorem. This definition is useful only insofar as it leads to interesting results.

\subsection{From dispersion relation to spectral dimension}

Factorize $P(s)$ into ``time'' and ``space'' contributions
\begin{equation} 
P(s) = \left[  \int {\d\omega \over2\pi} \; \exp\{-s\, \omega^2 \} \right] \times \left[ 
  \int {\d^d k\over(2\pi)^{d}} \; \exp\{-s\, f(k^2) \} \right].
\end{equation}
The first integral is elementary, so one has
\begin{equation}
P(s) =  \left[  {1\over \sqrt {4 \pi s}}  \right] \times \left[ \int  {\d^d k\over(2\pi)^{d+1}} \exp\{-s\,f(k^2) \} \right].
\end{equation}
Dropping irrelevant constants of proportionality (they would in any case drop out in the calculation of the spectral dimension) we see
\begin{equation}
P(s) \propto  {1 \over \sqrt s} \int  k^{d-1} \exp\{-s\,f(k^2) \} \d k.
\end{equation}
That is
\begin{equation}
\ln P(s) = -{1\over2} \ln s + \ln \int  k^{d-1} \exp\{-s\,f(k^2) \} \d k + C.
\end{equation}
(Note that $C$ will be used to denote a generic constant, when we re-use the symbol below it will not necessarily represent the \emph{same} constant.) 
This expression lets you calculate the \emph{exact} spectral dimension as
\begin{equation}
\label{exactds}
d_S(s) = 1 + 2s \; {  \int  f(k^2) \; k^{d-1} \; \exp\{-s\,f(k^2) \} \d k\over  \int k^{d-1}\;  \exp\{-s\,f(k^2)\; \} \d k }.
\end{equation}
Note that the leading ``1'' comes from the fact that the physical time part of the differential operator $D_E$ is trivial --- it is just $\partial_t^2$. All the complications come from the $d$ spatial dimensions.  
If we now write the dispersion relation as
\begin{equation}
\omega^2 = f(k^2) = \Omega(k)^2 ,
\end{equation}
then, still as an \emph{exact} result:
\begin{equation}
d_S(s) = 1 + 2s \; {  \int  \Omega(k)^2 \; k^{d-1} \; \exp\{-s  \Omega(k)^2 \} \d k \over  \int k^{d-1}\;  \exp\{-s  \Omega(k)^2\} \; \d k }\,.
\end{equation}
It is now useful to define a partition function 
\begin{equation}
\label{partf}
Z(s) =  \int k^{d-1}\;  \exp\{-s\,f(k^2)\, \} \; \d k =  \int k^{d-1}\;  \exp\{-s\,\Omega(k)^2\, \} \, \d k\,.
\end{equation}
Now by construction $Z(s) \propto P(s)$, so on the one hand it carries no extra information.  On the other hand it lets us 
 formally rewrite the spectral dimension as an ensemble average:
\begin{equation}
d_S(s) = 1-2s\; {\d \ln Z(s)\over\d s} = 1 + 2s \; \langle \Omega(k)^2 \rangle_s.
\end{equation}
Note that this partition function encodes relatively simple information concerning the dispersion relation of the specific degree of freedom under consideration. It is by no means the partition function of the entire system. 

This completes the first (and simplest) part of the programme: From any arbitrary dispersion relation $\omega = \Omega(k)$, and specified topological dimension $d$,  we have seen how to construct  suitable differential operators $D_L$ amd $D_E$ that encode this dispersion relation, and how to use the generalized diffusion process associated with the differential operator $D_E$ in order to define the corresponding spectral dimension $d_S(s)$.

\subsection{From spectral  dimension to dispersion relation}

The definitions given above allow one to determine the spectral dimension for a given dispersion relation. This is the usual case when one deals with a continuum theory and can linearize the field equations in order to determine the dispersion relation. In such cases the spectral dimension acts as an additional probe of the kinematics described by this dispersion relation (and potentially the geometry, see also section \ref{S:IIE2} below).

There are, however, cases where one can calculate the spectral dimension by other techniques and it then becomes useful to determine a corresponding (effective) dispersion relation. A typical example arises in discrete theories, such as CDT, where one can define and calculate the spectral dimension directly from a discretized diffusion process~\cite{CDT5,CDT6}. Clearly, in such theories the spectral dimension is not just an alternative to studying directly the propagator. Since the latter is simply not available, the former becomes a far more essential tool for obtaining information about the theory. Determining a corresponding (effective) dispersion relation could facilitate the comparison to some candidate continuum effective theory, see reference~\cite{CDT-Horava} for an example.

\subsubsection{Dispersion relation as inverse Laplace transform}

\emph{Formally} one can invert the process above and from the spectral dimension (as a function of $s$), plus the specified topological dimension,  reconstruct the dispersion relation. This is best done in two steps.
From
\begin{equation}
 {\d \ln Z(s)\over\d s} = - {d_S(s)-1\over 2 s},
\end{equation}
we have
\begin{equation}
\label{Zofs}
Z(s) = Z(s_0) \; \exp\left\{ -{1\over2} \int^s_{s_0} {d_S(\bar s)-1\over \bar s} \d \bar s  \right\}.
\end{equation}
That is, up so some constant of proportionality $Z(s_0)$, we can certainly reconstruct the partition function $Z(s)$ from the spectral dimension.
Can we take this any further? Can we reconstruct the dispersion relation $\Omega(k)$ from the spectral dimension $d_S(s)$?
After some mathematical manipulations one can bring equation~(\ref{partf}) in the form 
\begin{equation}
Z(s) = {1\over d} \int_0^\infty  \exp\{-s  \Omega^2\} \; {\d \{k(\Omega)^d\} \over\d[\Omega^2]} \; \d[\Omega^2]. 
\end{equation}
Performing an integration by parts yields
\begin{equation}
\int_0^\infty  \exp\{-s  \Omega^2\} \; \{k(\Omega)^d\} \; \d[\Omega^2] = {d\over s}  Z(s).
\end{equation}
This has the form of a Laplace transform, in the variable $\Omega^2$, of the function $k(\Omega)^d$. 

Implementing the inverse Laplace transform via a complex integration we have
\begin{equation}
k(\Omega)^d  = {1\over 2\pi i} \int_C   {d\over s} \;  Z(s)\;  e^{\Omega^2 s} \; \d s,
\end{equation}
where $C$ is an appropriate contour in the complex plane and $Z(s)$ is given by equation~(\ref{Zofs}). Therefore, in principle one can calculate the dispersion relation when the spectral dimension is analytically known on the complex plane as a function of $s$.

\subsubsection{Practical inversion techniques}

Though it settles an important issue of principle, the method of calculating the dispersion relation from the spectral dimension presented above is usually of little practical value. This is because, in either discrete or continuum theories, when one might be interested in determining a corresponding dispersion relation from a spectral dimension, the latter is typically not known analytically as a function of $s$. Hence, an algorithm that requires analytic continuation of the spectral dimension to complex values of $s$ is impractible. 
This problem can be circumvented in various ways:
\begin{itemize}
\item {\em Direct inversion via semi-analytic model building:} If one has somehow, (for instance, from some numerical simulation followed by some data-fitting), built a semi-analytic model for the spectral dimension $d_S(s)$, then integrating to find the partition function $Z(s)$ is typically easy. This semi-analytic model partition function $Z(s)$ can be analytically continued to the complex plane, and sometimes one is lucky --- one may encounter a model for the partition function for which the inverse Laplace transform is explicitly known.  Even when this works, we emphasise that one is analytically continuing the model, not the data. 

\item {\em Direct inversion via infinite differentiation:} There is a little-known method due to Post~\cite{Post}, see also Bryan~\cite{Bryan}, that allows for inversion of Laplace transforms by taking arbitrarily high derivatives.
Specifically, if $G(s)$ is the Laplace transform of $g(t)$ then
\begin{equation}
g(t) = \lim_{n\to\infty}  {(-1)^n\over n!} \; \left({n\over t}\right)^{n+1}  \; G^{(n)} \left({n\over t}\right).
\end{equation}
This algorithm may not always be practical, since one needs arbitrarily high derivatives. Even if not always practical, it again settles an important issue of principle --- --- knowledge of the spectral dimension $d_S(s)$ in principle allows one to reconstruct an equivalent dispersion relation $\Omega(k)$.

\item {\em Indirect  inversion via nonlinear regression:}
In contrast, a \emph{practical} algorithm for extracting (or rather, approximating) the dispersion relation is to build some theoretically appealing model for $\Omega(k)$ with several adjustable parameters, and to then use the technique of nonlinear regression to find a best fit within that class of dispersion relations. In a companion paper~\cite{CDT-Horava} we have applied this idea to the measured spectral dimension function arising in CDT models of quantum gravity.
\end{itemize}

\subsection{Some generalizations}

Let us now indicate some possible generalizations and extensions of this formalism.

\subsubsection{Beyond second order in physical time derivatives}

If one is (perhaps foolishly) willing to play with ghost fields then there is no need to restrict to second order derivatives in physical time. Consider the operator
\begin{equation}
D \Phi \equiv  f(- \partial_t^2,-\nabla^2) \; \Phi.
\end{equation}
The relevant dispersion relation is now only \emph{implicitly} defined by
\begin{equation}
\label{fdisp}
f(\omega^2,k^2) = 0.
\end{equation}
On the other hand the heat kernel becomes
\begin{equation}
\rho(x,x',s) = \int {\d^d k \; \d \omega \over(2\pi)^{d+1}}   \exp\{ i [\vec k \cdot (\vec x-\vec x') +\omega(t-t')] \} \exp\{-s[f(\omega^2, k^2) ]\},
\end{equation}
and the diagonal part of the heat kernel specializes to
\begin{equation}
P(s) = \rho(x,x,s) =  \int {\d^d k \; \d \omega \over(2\pi)^{d+1}} \exp\{-s \, f(\omega^2,k^2)\}.
\end{equation}
So the spectral dimension is 
\begin{equation}
d_S(s) = 2 s \;
{\int f(\omega^2,k^2)  \exp\{-s \, f(\omega^2,k^2)\} \;\d \omega \; k^{d-1} \; \d k \over
\int \exp\{-s \, f(\omega^2,k^2)\} \; \d \omega \; k^{d-1} \; \d k } = 2s \langle f(\omega^2,k^2) \rangle_s 
\end{equation}
and depends on the whole of the function $f(\omega^2,k^2)$, while the dispersion relation is given by the solution of equation~(\ref{fdisp}) and uses only limited information about the function $f(\omega^2,k^2)$.

If the differential operator can be put in the form
\begin{equation}
D \Phi \equiv  \left\{ f_1(- \partial_t^2) + f_2(-\nabla^2) \right\} \Phi,
\end{equation}
then the space and time contributions to the spectral dimension become separable and one has
\begin{equation}
d_S(s) =  2s \; {  \int  f_1(\omega^2) \, \exp\{-s\,f_1(\omega^2) \}\, \d \omega\over  \int \exp\{-s\,f_1(\omega^2)\, \} \,\d \omega }
 + 2s \; {  \int  f_2(k^2) \; k^{d-1} \; \exp\{-s\,f_2(k^2) \} \, \d k\over  \int k^{d-1}\;  \exp\{-s\,f_2(k^2)\, \}\, \d k }.
\end{equation}
Clearly, when $f_1(- \partial_t^2)=- \partial_t^2$, one recovers the result given in equation~(\ref{exactds}).

\subsubsection{Curved spacetime}
\label{S:IIE2}

A perhaps more interesting question is to ask what happens if spacetime is allowed to be curved, in addition to having a nontrivial dispersion relation. Extending the definition of the spectral dimension to dispersion relations describing propagators in a {\em given} curved background is actually rather straightforward: $D_E$ in equation~(\ref{diffusion}) will be a differential operator in curved space, but the solution of this equation can still be used to calculate the return probability. The only subtlety is that when one has to deal with spaces which do not respect translational invariance, the return probability would be space dependent, $P(x, s)$. Thus, so long as one does not want the spectral dimension to depend on the spacetime point, one has to actually use the average return probability in the definition of equation~(\ref{dsdef}) (as was mentioned already in the the introduction). An Arnowitt--Desser--Misner decomposition provides a natural setting for constructing differential operators which include curvature and at the same time lead to nontrivial dispersion relations even around maximally symmetric spaces.

\subsection{Space versus spacetime}

The $d_S(s)$ we have so far been considering is the spectral dimension of \emph{spacetime}. If one is only interested in the spectral dimension of \emph{space}, (as is usual for most classical [non-relativistic] physicists), then the appropriate thing to do is to simply drop the $t$ coordinate and define
\begin{equation}
D_\mathrm{space} = - f(-\nabla^2).
\end{equation}
One would then consider the diffusion process
\begin{equation}
\left\{ {\partial\over\partial s}  - f(-\nabla^2) \right\} \rho(x,x', s) = 0; \qquad   \rho(x,x',0) = \delta^{d}(x-x').
\end{equation}
In this case (now without the leading ``1'') we have
\begin{equation}
d_{S,\mathrm{space}}(s) =  2s \; {  \int  f(k^2) \; k^{d-1} \; \exp\{-s\,f(k^2) \} \d k\over  \int k^{d-1}\;  \exp\{-s\,f(k^2)\; \} \d k },
\end{equation}
or equivalently
\begin{equation}
d_{S,\mathrm{space}}(s) =  2s \; {  \int  \Omega(k)^2 \; k^{d-1} \; \exp\{-s  \Omega(k)^2 \} \d k \over  \int k^{d-1}\;  \exp\{-s  \Omega(k)^2\} \; \d k } = 2s \; \langle \Omega(k)^2 \rangle_s\,.
\end{equation}
Note that (assuming physical time only shows up in the simple form $\partial_t^2$) the only difference between the spectral dimension of spacetime, and that of space itself, is the leading ``1''. 
Furthermore, since $t$ has been eliminated from the formalism, the diffusion time $s$ can in this situation be identified with physical Newtonian time.

\section{Calculational techniques}

We shall now explore some calculational techniques for determining the spectral dimension $d_S(s)$. In simple situations we can obtain exact analytic results, but more typically one has to resort to some form of approximation. We shall show that making use of asymptotic expansions can be fruitful in both the infrared and ultraviolet, and that saddle point techniques can also be employed --- typically over the whole range of interest.

In the following discussion we will also present several examples. For simplicity we will largely focus on polynomial dispersion relations, where
\begin{equation}
\omega^2 = c_w k^{2w} + c_{w+1}  k^{2w+2} + \dots + c_{z-1} k^{2z-2} + c_z k^{2z}.
\end{equation}
Here $w\leq z$ are integers potentially unrelated to $d$. As additional examples (better dealt with using asymptotic or saddle point techniques) we will also use rational polynomial dispersion relations of the form
\begin{equation}
\omega^2 =   {p(k^2)\over q(k^2)},
\end{equation}
where $p(k^2)$ and $q(k^2)$ are polynomials in $k^2$, and surface-wave dispersion relations that sometimes include trigonometric functions.

The choice of these specific types of dispersion relations as examples is motivated more by mathematical simplicity and their demonstrative power, than by their origin or genericity. Nevertheless, all of them are common (to different extents). Polynomial dispersion relations are typical in (truncated) effective field theories that exhibit Lorentz violations. They also appear in projectable Ho\v rava  gravity~\cite{Horava1, SVW1, SVW2, SVW3}. (Projectability requires that the lapse function is a function of time only, $N=N(t)$.) In the latter case, the dispersion relation is  actually polynomial in $k^2$ with the highest power being $(k^2)^d$, that is, 
\begin{equation}
\omega^2 = c_1 k^2 + c_2 k^4 + \dots + c_d k^{2d}.
\end{equation}
 and with particularly pleasant renormalizability properties being associated with the choice $z=d$ \cite{Horava1, LSB, LSB2}. Rational polynomial dispersion relations are generic in more general, non-projectable Ho\v rava gravity \cite{Blas, 2+1-in-preparation}. (See reference~\cite{Sotiriou:2010wn} for a brief review on Ho\v rava gravity.) Finally, gravity-driven surface waves have dispersion relations that include trigonometric functions. In any case, the techniques we present here are applicable to more general dispersion relations.

\subsection{Exact results}

Useful exact results seem to be limited to special cases of the two-term dispersion relation
\begin{equation}
\omega^2 = c_w k^{2w} + c_z k^{2z}. 
\end{equation}
More complicated dispersion relations, or even the general case above, typically require one to invoke some sort of series expansion or approximation technique.

\subsubsection{Bogoliubov dispersion relation}
The so-called Bogoliubov dispersion relation can conveniently be put in the form
\begin{equation}
f(k^2) =  k^2+{k^4\over 2^d \; K^2};  \qquad \Omega(k) =  k \sqrt{1+{k^2\over 2^d \; K^2}}.
\end{equation}
(The factor of $2^d$ is there just to minimize irrelevant numerical factors in the results below.)
\paragraph{Two space dimensions:}
The partition function for the Bogoliubov dispersion relation can be explicitly evaluated in terms of the error function
\begin{eqnarray}
Z(s) &=& \int  k \exp\left\{-s k^2 \left(1+{k^2\over 4K^2}\right) \right\} \d k 
\\
&=&  {K\over2 } \sqrt{\pi\over s}  \exp\left({s K^2}\right) \left[ 1-\mathrm{erf}\left(\sqrt{s K^2}\right) \right].
\end{eqnarray}
The spectral dimension is then explicitly evaluable as
\begin{equation}
d_S(s) = 2 - 2s K^2 + 2\sqrt{s K^2\over\pi} \; {e^{-sK^2}\over1-\mathrm{erf}(\sqrt{s K^2})}.
\end{equation}
Note the limits
\begin{equation}
d_S(s\to 0) =  2; \qquad d_S(s\to \infty) =  3.
\end{equation}

\paragraph{Three space dimensions:}
In this situation the partition function for the Bogoliubov dispersion relation can be explicitly evaluated in terms of  Bessel functions:
\begin{eqnarray}
Z(s) &=& \int  k^2 \exp\left\{-s k^2 \left(1+{k^2\over 8K^2}\right) \right\} \d k 
\\
&=&  \sqrt{2} {K^3} \exp\left({s K^2}\right) \left[ K_{3/4}\left( {s K^2}\right) - K_{1/4}\left({s K^2}\right) \right].
\end{eqnarray}
The spectral dimension is then explicitly evaluable as
\begin{equation}
d_S(s) = 2 - 4s K^2 - {1\over2} 
\left[ {K_{3/4}\left( {s K^2}\right) + K_{1/4}\left({s K^2}\right)\over K_{3/4}\left( {s K^2}\right) - K_{1/4}\left({s K^2}\right) } \right] .
\end{equation}
Note the limits
\begin{equation}
d_S(s\to 0) =  {5\over2}; \qquad d_S(s\to \infty) =  4.
\end{equation}

\subsubsection{Quadratic plus sextic dispersion relation}
Consider the specific dispersion relation
\begin{equation}
f(k^2) =  k^2 \left(1+{k^4\over27K^4}\right) = k^2 + {k^6\over27 K^4};  \qquad \Omega(k) =  k \sqrt{1+{k^4\over27K^4}}.
\end{equation}
(The ``27'' is there just to minimize irrelevant numerical factors in the results below.)
This dispersion relation is potentially interesting as can serve as a good approximation for both the high-momentum and low-momentum limit of Ho\v rava gravity~\cite{Horava} (it interpolates between low-momentum Lorentz invariance and a $z=3$ Lifshitz point at large momentum). 
It is not, however, the exact dispersion relation of any of the versions of the model in 3+1 dimensions, (there is no $k^4$ term, and it is not a rational polynomial).

The partition function can be explicitly evaluated in terms of Bessel functions
\begin{eqnarray}
Z(s) 
&=& \int  k^2 \exp\left\{-s  k^2 \left(1+{k^4\over27K^4}\right) \right\} \d k 
\\
&=& - {3\over4} \pi^{3/2} K^4 \sqrt{s} \left[ J_{1/3} \left( {s  K^2} \right) J_{-2/3}\left( {s K^2}\right) 
 + Y_{1/3}\left({s K^2}\right) Y_{-2/3}\left({s K^2}\right)  \right].
 \nonumber
\end{eqnarray}
The spectral dimension is then explicitly evaluable as
\begin{equation}
d_S(s) = 2 + 2 s K^2 
\left[{  
J_{1/3} \left( {s  K^2} \right)^2- J_{-2/3}\left( {s K^2}\right)^2 +  Y_{1/3}\left({s K^2}\right)^2  - Y_{-2/3}\left({s K^2}\right)^2
\over
J_{1/3} \left( {s  K^2} \right) J_{-2/3}\left( {s K^2}\right)  + Y_{1/3}\left({s K^2}\right) Y_{-2/3}\left({s K^2}\right) 
 }\right].
\end{equation}
Note the limits
\begin{equation}
d_S(s\to 0) =  2; \qquad d_S(s\to \infty) =  4.
\end{equation}

\subsubsection{Quartic plus sextic dispersion relation}
As a final exact example, albeit one where the results are too messy to actually explicitly write down, in (3+1) dimensions consider the dispersion relation 
\begin{equation}
f(k^2) =  3 k^4 + {2 k^6\over K^2}.
\end{equation}
The numerical factors are again chosen to keep expressions as simple as possible.
The partition function can then be evaluated as a sum of generalized hyper-geometric functions, specifically as a sum of three $_2F_2(-sK^4)$ hyper-geometric functions. 
We have
\begin{eqnarray}
Z(s) 
&=& \int  k^2 \exp\left\{-s \left(3 k^4+{2k^6\over K^2}\right) \right\} \d k \nonumber
\\
&=&  {1\over12} \sqrt{2\pi K^2\over s} \Bigg[ \; {}_2F_2{}_{\left([1/4,3/4],[1/3,2/3]\right)}(-sK^4)
 \nonumber\\
 &&
 \qquad
-{\Gamma({1\over6})\over4\sqrt{\pi} }\;  (2sK^4)^{1/3} \;   {}_2F_2{}_{\left([7/12,13/12],[2/3,4/3]\right)}(-sK^4)
 \nonumber\\
 &&
 \qquad
 +{15\Gamma({5\over6})\over 16 \sqrt{\pi}}\;  (2sK^4)^{2/3} \;   {}_2F_2{}_{\left([11/12,17/12],[4/3,5/3]\right)} (-sK^4)
 \Bigg]. \;\;
\end{eqnarray}
The resulting spectral dimension is too messy to be worth writing down explicitly, but the limits are straightforward to evaluate:
\begin{equation}
d_S(s\to 0) =  2; \qquad d_S(s\to \infty) =  {5\over2}.
\end{equation}

\subsubsection{Summary of exact results}
It should be obvious by now that even relatively simple dispersion relations, such as 
\begin{equation}
\omega^2 = c_w k^{2w} + c_z k^{2z}, 
\end{equation}
lead to rather complicated expression for the spectral dimension. Therefore, it is interesting to develop techniques that could allow us to determine the spectral dimension to some required accuracy based on a suitable approximation. This will be the aim of the in the following sections.

Apart from demonstrating the difficulty of determining the spectral dimension for a given dispersion relation, the exact results presented above provide also some insight into how the spectral dimension changes with the length scale. Given that $s$ is a (fictitious) diffusion time, small values of $s$ probe small length scales and large values of $s$ probe large scales. This is the reason  why we paid particular attention to the limits $s\to 0$ and $s\to \infty$. For dispersion relations of the form given above and 
whenever we have been able to explicitly carry out the exact calculation we have found that
\begin{equation}
d_S(s\to 0) =  1 + {d\over z}; \qquad d_S(s\to \infty) =  1 + {d\over w},
\end{equation}
in $(d+1)$ topological dimensions.

Note that, from a field theory point of view it is customary to identify the small length scale limit with $k\to \infty$ and the large length scale limit with $k\to 0$. Indeed, it is intuitive that there is generically correspondence between $s\to 0$ and $k\to \infty$ or between $s\to \infty$ and $k\to 0$. In fact, if one is interested in calculating only these limits, it is much easier to suitably truncate the dispersion relation for small or large $k$ and then directly calculate the spectral dimension in these limits (as was done in reference~\cite{Horava}).

It is worth mentioning that the correspondence between the two limits (in $s$ and $k$) can be readily seen from the end result of the three examples we gave above. Notice that in all three cases the spectral dimension is actually a function of $s K^n$, where $K$ is the energy scale suppressing the higher order momentum terms in the dispersion relation (and $n$ is the exponent of the lower order term based on dimensional analysis). Hence, clearly the limit $s\to 0$ ($s\to \infty$) will give no different results than the limit $K\to 0$ ($K\to \infty$). However, given the form of dispersion relations we are considering, driving $K$ to infinity is no different than taking $k\to 0$, and driving $K$ to zero is not different than taking $k\to \infty$.

The general pattern relating the small/large length scales (ultraviolet/infrared limits) of the spectral dimension suggested by our analytic examples will be be investigated in the next two sections using both asymptotic techniques and saddle point methods.

\subsection{Asymptotic expansions}
Since obtaining explicit analytic results for the spectral dimension seems to be limited to very simple dispersion relations, it is worth developing perturbative expansions that could help us determine the former from the latter to some desired accuracy in a regime of interest. Specifically, if one is interested in studying the very small length scale (ultraviolet) or the very large length scale (infrared) behaviour of the spectral dimension, then 
asymptotic techniques are particularly useful. These are based on the idea of dividing the exponential appearing in the partition function, $e^{-s f(k^2)}$, into a dominant piece plus perturbative corrections. Exactly how one does the division will depend on which of the two asymptotic regimes one is interested in probing.

Consider a dispersion relation of the form
\begin{equation}
\omega^2 = f(k^2) = c_w k^{2w} + c_{w+1}  k^{2w+2} + \dots + c_{z-1} k^{2z-2} + c_z k^{2z},
\end{equation}
with $1\leq w < z$. Then one can certainly write both
\begin{equation}
\label{E:w}
\exp[-s f(k^2)] = \exp(-s c_w k^{2w}) \left\{ 1 - s c_{w+1} k^{2w+2} + \dots \right\},
\end{equation}
and
\begin{equation}
\label{E:z}
\exp[-s f(k^2)] = \exp(-s c_z k^{2z}) \left\{ 1 - s c_{z-1} k^{2z-2} +  \dots \right\},
\end{equation}
by performing  expansions around $k=0$ and $s=0$ respectively. The series in braces consist of various positive powers of $s$ and $k$ and are guaranteed convergent for all values of $s$ and $k$.  The sub-leading terms appearing in equation (\ref{E:w}) are all of the form $s^p k^{2q}$ with $q\geq p(w+1)$. Similarly sub-leading terms appearing in equation (\ref{E:z}) are all of the form $s^p k^{2q}$ with $q\leq p(z-1)$.

These two expansions can be very useful for studying the infrared and ultraviolet behaviour of the spectral dimension. Moreover, they are not actually limited to polynomial dispersion relations. Consider instead a general dispersion relation $\omega^2=f(k^2)$. If $f(\cdot)$ does not have a pole at $k=0$, then one can always write
\begin{equation}
\omega^2 = f(k^2) = c_w k^{2w} + c_{w+1}  k^{2w+2} + \dots,
\end{equation} 
which is sufficient to obtain equation~(\ref{E:w}). On the other hand, as long as $f(\cdot)$ has a pole of finite order at $k\to \infty$, one can always write the dispersion relation for very large $k$ as a Taylor series of descending powers of $k$, {\em i.e.}
\begin{equation}
\omega^2 = f(k^2) =c_z k^{2z}+ c_{z-1} k^{2z-2} +\dots.
\end{equation}
Again, this is sufficient for obtaining equation~(\ref{E:z}). Both requirements are fulfilled by reasonable and phenomenologically interesting dispersion relations.

We are now ready to proceed to suitable ultraviolet and infrared perturbative expansions. Before doing so, it is worth mentioning the following.
We have restricted our discussion so far to dispersion relations that respect parity invariance. This is a very well motivated simplifying assumption: The lowest order term that violates parity ($\omega^2\propto k$) is important at low energies and would, therefore, lead to severe disagreement with observations. However, one could relax parity invariance and introduce some other symmetry that would prohibit the presence of lower-order but not higher-order parity violating terms (one example seems to be Ho\v rava's detailed balance condition \cite{Horava1}). In any case, even though we will retain parity invariance for the sake of brevity, it is not crucial for our approach. After deriving the asymptotic expansions, we will comment on how they would be modified if parity invariance were to be abandoned.

\def\O{\mathcal{O}}

\subsubsection{Deep ultraviolet (generalized Seeley--DeWitt expansion)}
We now argue that there is an appropriate generalization of the usual Seeley--De~Witt expansion beyond ordinary second-order differential operators. If $f(-\nabla^2)$ is polynomial and contains terms up to order $\nabla^{2z}$, 
 then there will exist an asymptotic expansion (in terms of \emph{positive fractional} powers of $s$) of the form
\begin{equation}
\label{sde}
 Z(s)=   {C \over s^{d/(2z)}} \left\{ \sum_{n=0}^N   a_n s^{n /z} + \O\left(s^{(N+1)/z}\right) \right\},
 \end{equation}
where we normalize to $a_0=1$, (and $C$ is for our current purposes uninteresting).  The usual Seeley--De~Witt expansion corresponds to $z=1$. Note that we have already peeled off physical time and are only dealing with the spatial part of the heat kernel. To convince yourself that this expansion holds, proceed as follows:
Focus on the highest-derivative piece. A straightforward computation (using the definition if the $\Gamma$ function) yields
\begin{equation}
Z_{(-\nabla^2)^z}(s)=    \int k^{d-1} e^{-s k^{2z} } \d k = {\Gamma({d\over2z})\over2z}  s^{-d/(2z)}.
\end{equation}
Note that  the sub-leading terms are all of the form $s^p k^{2q}$ and so will contribute 
\begin{equation}
 \int k^{d-1} \{ s^p k^{2q} \} \exp(-s k^{2z} ) \; \d k
 \propto s^{-d/(2z)} \cdot s^p\cdot s^{-q/z}.
\end{equation} 
So in particular,  for the first sub-leading term in equation (\ref{E:z}) which is proportional to $s k^{2z-2}$, we have
\begin{equation}
 \int k^{d-1} \{ s k^{2z-2} \} \exp(-s k^{2z} ) \; \d k
 \propto   s^{-d/(2z)} \cdot s \cdot s^{-1}\cdot s^{+1/z} =  s^{-d/(2z)} \cdot s^{+1/z}.
\end{equation} 
Then, at the very least $Z(s)$ will contain the terms
\begin{equation}
Z(s) = C\{ a_0 s^{-d/(2z)} +  a_1  s^{-d/(2z)} \cdot s^{1/z} + \dots \}.
 \end{equation}
 Higher order terms $s^p k^{2q}$ contribute to the partition function with relative strength $s^p \cdot s^{-q/z}$, that is $s^{(pz-q)/z}$.  But $pz-q \geq p$ is a positive integer, so these are all positive powers $s^{n/z}$ of $s^{1/z}$.  
That is, this series for $Z(s)$  (which can at best be an asymptotic series) continues with increasing powers of $(s^{1/z})^n$. This implies that $Z(s)$ is indeed given by equation~(\ref{sde})
 as claimed.

Now, turning to the spectral dimension, one can straightforwardly derive 
\begin{equation}
\label{dssde}
d_S(s) = 1 + {d\over z} -   {2 a_1\over z} s^{1/z} +  \O(s^{2/z}).
\end{equation}
So, in the deep ultraviolet ($s\to 0$) the spectral dimension flows generically (modulo our assumptions about parity invariance) to 
\begin{equation}
d_S(s) \to 1 + {d\over z},
\end{equation}
with a prescribed rate of $\O(s^{1/z})$, where $2z$ is the order of the pole of the dispersion relations as $k\to \infty$ (and $d$ the topological spatial dimension).
We are (for current purposes) not particularly interested in the specific value of the coefficient $a_1$, though it can be calculated without any particular difficulty. 
When we normalize to $a_0=1$ a brief calculation yields
\begin{equation}
a_1 = - {c_{z-1}\over (c_z)^{(z-1)/z} } \; {\Gamma({d-2\over2z} + 1)\over \Gamma({d\over2z})}.
\end{equation}
 This quantity $a_1$ is not, in any sense, universal like the usual Seeley--De~Witt coefficients --- $a_1$ will depend explicitly on the coefficients in the dispersion relation $f(k^2)$. 
Had we allowed for parity violations the only qualitative modification in the spectral dimension as given by equation~(\ref{dssde}) would be that $z$ would be allowed to take half-integer values. 

\subsubsection{Deep infrared (generalized Feynman expansion)}
Turning to the infrared, we now argue that there is an appropriate ``inverted'' generalization of the Seeley--De~Witt expansion beyond ordinary second-order differential operators. (This is very close in spirit to the usual Feynman diagram expansion around a Gaussian integral.) Specifically, if $w=1$ so that  the low-momentum dispersion relation starts off as $f(k^2) = k^2 +...$, we assert the existence of an asymptotic expansion (now in \emph{inverse integer} powers of $s$) of the form
\begin{equation}
 Z(s) =   {C \over s^{d/2}} \left\{ \sum_{n=0}^N   \tilde a_n s^{-n} + \O\left(s^{-(N+1)}\right) \right\},
 \end{equation}
where we can normalize to $\tilde a_0=1$.  Even if we do not have Lorentz invariance in the low energy limit ($w>1$), we can still make the general assertion
\begin{equation}
\label{gfe}
 Z(s) =   {C \over s^{d/2w}} \left\{ \sum_{n=0}^N   \tilde a_n s^{-n/w} + \O\left(s^{-(N+1)/w}\right) \right\}.
 \end{equation}

To convince yourself that this is true, proceed as follows (many of the technical steps are very similar to those for the ultraviolet limit and some details will be suppressed).
Focus on the lowest-derivative piece,  for which we have
\begin{equation}
 Z_{(-\nabla^2)^w}(s) =    \int k^{d-1} e^{-s k^{2w} } \d k \propto s^{-d/2w}.
\end{equation}
Furthermore, note that all the sub-leading terms are of the form $s^p k^{2q}$, and so by a minor variant of the previous argument they contribute quantities of relative magnitude $s^p \cdot s^{-q/w}$ to the partition function $Z(s)$. 
But the first sub-leading term in equation (\ref{E:w}) is proportional to $s k^{2w+2}$, so at the very least $Z(s)$ will contain the terms
\begin{equation}
 Z(s)  = C \{ a_0 s^{-d/(2w)} + s \cdot a_1  s^{-d/(2w)} \cdot s^{-(w+1)/w} + \dots \}.
 \end{equation}

Higher order terms contribute with relative strength $s^p \cdot s^{-q/w} = s^{(pw-q)/w}$, but from equation (\ref{E:w})  we know $q>p(w+1)$ so $pq-q<-p$ is a negative integer. So higher order terms contribute with relative strength $s^{-n/w}$, that is, with positive integer powers of $s^{-1/w}$. 
Thus,  $Z(s)$ is indeed given by equation~(\ref{gfe})
 as claimed. The corresponding spectral dimension is then 
 \begin{equation}
d_S(s) = 1 + {d\over w}  +   {2 \tilde a_1} s^{-1/w} +  \O(s^{-2/w}).
\end{equation}
Thus deep in the infrared ($s\to\infty$), as long as the low-momentum dispersion relation is of the form $f(k^2) = k^2 + k^4+ ...$,  (that is $w=1$), the spectral dimension will flow to $d+1$, the topological  dimension of spacetime, at a rate $\O(1/s)$. More generally, it will flow to $1+d/w$ at a rate of $\O(s^{-1/w})$. As before, had we given up parity invariance, $w$ would be allowed to take half-integer values.
Finally, $\tilde a_1$ is not in any sense universal and can  be calculated without difficulty.  Normalizing to $\tilde a_0=1$, a brief calculation yields
\begin{equation}
\tilde a_1 = - {c_{w+1}\over (c_w)^{(w+1)/w}}\;  {\Gamma({d+2\over 2w} +1) \over\Gamma({d\over2w})}.
\end{equation}

\subsubsection{Example: Rational polynomial dispersion relation}
\label{S:rat-poly}

As previously mentioned, non-projectable Ho\v rava gravity models lead to dispersion relations that are rational ratios of polynomials in $k^2$ \cite{Blas,2+1-in-preparation} so that
\begin{equation}
\omega^2 =   {p(k^2)\over q(k^2)}\,.
\end{equation}
Since $p(k^2)$ and $q(k^2)$ are both polynomials in $k^2$ we can argue that
\begin{equation}
 {p(k^2)\over q(k^2)} \to c_z k^{2z} + \O(k^{2z-2}) \quad \hbox{as} \quad k \to \infty\,. 
\end{equation}
Then we have
\begin{equation}
f(k^2) = c_z k^{2z} + \O(k^{2z-2}) \quad \hbox{as} \quad k\to\infty\,.
\end{equation}
Therefore for the spectral dimension we have
\begin{equation}
d_S = 1 + {d\over z} + \O(s^{1/z}).
\end{equation}
A similar procedure works in the infrared.  At low $k$ any rational ratio $p(k^2)/q(k^2)$ can be expanded as a formally infinite power series in  $k^2$.  Ignoring issues of convergence, we can then still apply the generalized Feynman expansion to argue for behaviour of the form
\begin{equation}
d_S = 1 + {d\over w} + \O(s^{-1}),
\end{equation}
where we define $w$ by $p(k^2)/q(k^2) \to k^{2w} + \O(k^{2w+2})$ as $k\to 0$.

\subsection{Saddle point techniques}

We now have good control of both asymptotic limits, which (where they overlap) are in complete agreement with the exact results obtained previously. However, in the intermediate regime neither asymptotic expansion need be reliable --- neither generalized Seeley--De~Witt nor generalized Feynman expansions are guaranteed to be useful. Fortunately, in this situation saddle point techniques can provide significant and useful approximate information. 

Consider any integral of the form
\begin{equation}
J = \int_{-\infty}^{+\infty} \exp[h(x)] \; \d x.
\end{equation}
The saddle point approximation consists of first locating the maximum of the argument $h(x)$ by solving
\begin{equation}
h'(x_*) =0,
\end{equation}
and then approximating
\begin{eqnarray}
J &\approx&  \int_{-\infty}^{+\infty} \exp\left[h(x_*) + {1\over2} h''(x_*) (x-x_*)^2 \right]  \; \d x
\\
&=&   \exp\left[h(x_*) \right] \int_{-\infty}^{+\infty} \exp\left[ {1\over2} h''(x_*) (x-x_*)^2 \right]  \; \d x
\\
&=&  \exp\left[h(x_*) \right] \times  \sqrt{2\pi \over -h''(x_*)}.
\end{eqnarray}
For most purposes (especially when looking at ratios of integrals)  it is sufficient to use the even more brutal approximation
\begin{equation}
\int_{-\infty}^{+\infty} \exp[h(x)] \; \d x \approx \exp\left[h(x_*)\right].
\end{equation}
See for example reference~\cite{compact}, where similar considerations are applied to the stationary phase approximation. 

\subsubsection{General framework} 
Taking the change of variable $u = \ln k$ we can express the partition function as 
\begin{eqnarray}
Z(s) 
&=&  \int\exp\{u d  -s\,  f(e^{2u}) \}  \;\d u.
\end{eqnarray}
Note that
\begin{equation}
\{u d  -s\,  f(e^{2u}) \}' =  d - 2 s \,  f'(e^{2u})  e^{2u},
\end{equation}
and so the location of the saddle point is at
 \begin{equation}
 \label{spc}
 d = 2 s f'(e^{2u})  e^{2u} =  2 s f'(k_*^2) k_*^2,
 \end{equation}
 which implicitly defines $k_*(s)$. 
 We can rewrite this equation in terms of the phase and group velocities as
 \begin{equation}
 d = 2 s \, {\Omega_*\over k_*} {\d \Omega\over\d k_*} k_*^2 = 2 s \, v_\mathrm{phase}(k_*) \; v_\mathrm{group}(k_*) \; k_*^2,
 \end{equation}
 implying that the saddle point estimate of the diffusion time can be specified as a function of the wavenumber
 \begin{equation}
 \label{E:s}
 s(k_*) =  {d\over2  v_\mathrm{phase}(k_*) \; v_\mathrm{group}(k_*) \; k_*^2}.
 \end{equation}
 We now see
\begin{equation}
 Z(s) \approx   \exp[\{u d  -s\,  f(e^{2u}) \}_*]  =  k_*^d \; \exp\{-s\,  f(k_*^2) \}.
  \end{equation}
Therefore, in view of the saddle point condition in equation~(\ref{spc}), the leading term in the saddle point approximation yields for the spectral dimension
 \begin{equation}
d_S(s) \approx 1 + 2s  f(k_*^2)  = 1 + {d \; f(k_*^2)\over f'(k_*^2) k_*^2}.
\end{equation}
That is, expressing $s$ in terms of $k_*$,
\begin{equation}
d_S(k_*) = 1 + d\;  {v_\mathrm{phase}(k_*)\over v_\mathrm{group}(k_*)} + ...
\end{equation}
\begin{equation}
 \label{E:s2}
 s(k_*) =  {d\over2  v_\mathrm{phase}(k_*) \; v_\mathrm{group}(k_*) \; k_*^2}.
 \end{equation}
If desired, the curve $d_S(s)$ can be explicitly constructed by parametrically plotting the pair $\{ d_S(k_*),  s(k_*) \}$.

To once more verify our asymptotic results, let us now consider the situation
\begin{equation}
\Omega(k) \sim k^w \quad \hbox{as} \quad k \to 0; \qquad \hbox{and}\qquad \Omega(k) \sim k^z \quad \hbox{as} \quad k \to \infty. 
\end{equation}
We then have
\begin{equation}
s(k_*) \sim {d\over 2w k_*^{2w}}\to\infty \quad \hbox{as} \quad k_* \to 0; \qquad \hbox{and}\qquad s(k_*) \sim {d\over 2z k_*^{2z}}\to 0 \quad \hbox{as} \quad k_* \to \infty\,, 
\end{equation}
as expected. Additionally
\begin{equation}
d_S(k_*) \approx 1 + {d\over w} \quad \hbox{as} \quad k_* \to 0; \qquad \hbox{and}\qquad d_S(k_*) \approx 1 + {d\over z} \quad \hbox{as} \quad k_* \to \infty\,, 
\end{equation}
but now there is no requirement that $w$ and $z$ be integers. So, the infrared and ultraviolet limits for $d_S(s)$ that we first encountered in several exact examples, and then verified for general polynomial dispersion relations using asymptotic expansions, are now seen to have even more generality within the framework of the saddle point approximation.

\subsubsection{Example: Polynomial dispersion relation}

Suppose we have a polynomial dispersion relation
\begin{equation}
\Omega(k)^2  = \sum_{a=w}^z c_a k^{2a}.
\end{equation}
Then
\begin{equation}
v_\mathrm{group}(k) = {\sum_a  a c_a k^{2a-1}\over \sqrt{ \sum_a c_a k^{2a} }}; 
\qquad 
v_\mathrm{phase}(k) = { \sqrt{ \sum_a c_a k^{2a} }\over k};
\end{equation}
so the saddle point approximation implies
\begin{equation}
d_S(k_*) \approx 1 + d \; { \sum_a c_a k_*^{2a} \over \sum_a  a c_a k_*^{2a} }; 
\qquad 
s(k_*) = {d \over 2\sum_a  a c_a k_*^{2a} }.
\end{equation}
Then:
\begin{equation}
d_S(s\to \infty) = d_S(k_*\to 0) \approx 1 + {d\over w}; 
\quad
d_S(s\to 0) = d_S(k_*\to \infty) \approx 1+ {d\over z};
\end{equation}
which agrees with our explicit limits via the asymptotic analyses --- but now also gives us approximate information at intermediate regimes.

\subsubsection{Example: Two-term dispersion relation}
It is worthwhile to explicitly consider the simple two-term dispersion relation
\begin{equation}
\Omega(k)^2 = c_w k^{2w} + c_z k^{2z},
\end{equation}
since then we have particularly simple formulae
\begin{equation}
d(k_*) \approx 1 + d\; {c_w k_*^{2w} + c_z k_*^{2z}\over w c_w k_*^{2w} + z c_z k_*^{2z}};
\qquad
s(k_*) = {d\over  2( w c_w k_*^{2w} + z c_z k_*^{2z})}.
\end{equation}

\subsubsection{Example: Bogoliubov dispersion relation}
Consider the specific case of the Bogoliubov dispersion relation written in the form
\begin{equation}
\Omega(k)^2 = k^2 \left(1 + {k^2\over4K^2}\right).
\end{equation}
Then the saddle point approximation yields
\begin{equation}
d(k_*) \approx 1+ d\; {1+{k_*^2\over4K^2}\over1+{k_*^2\over2K^2}};
\qquad
s(k_*) = {d\over  k_*^2} \; {1\over 2 +  k_*^2/K^2}.
\end{equation}
In this particular case we can explicitly invert $s(k_*)$ to find $k_*(s)$:
\begin{equation}
k_*^2(s)  = K^2 \left(\sqrt{{d\over s  K^2} + 1} - 1\right).
\end{equation}
Thence
\begin{equation}
d_S(s) \approx 1 +  {d\over2} \; {3 s  K + \sqrt{ s(s K^2+d)} \over s K + \sqrt{ s(s K^2+d)} }.
\end{equation}
That is
\begin{equation}
d_S(s) \approx 1+ {d\over2} \; {3 + \sqrt{ 1+d/(s  K^2)} \over 1 + \sqrt{ 1+d/(s  K^2)} }.
\end{equation}
This compact and explicit (though approximate) formula has the correct limits. ($d_S(s\to\infty)= d+1$; while $d_s(s\to0)= 1+ d/2$.)
This result can in $2+1$ dimensions be explicitly compared to the exact analytic result obtained via the Gaussian error function. 

\subsubsection{Example: Surface waves}
Take a two-dimensional liquid-gas interface, (topological dimension $d=2$), and consider the surface waves. For example, finite-depth ocean waves ignoring surface tension. Then the dispersion relation is well known to be
\begin{equation}
\Omega(k)^2 = g k \tanh (k h),
\end{equation}
where $h$ is the depth of the ocean.
Since this an intrinsically non-relativistic system, as we have previously argued it is best to simply consider the spectral dimension of space (rather than spacetime). 
The saddle point approximation leads to
\begin{equation}
d_{S,\mathrm{space}}(k_*) \approx { 4\sinh(2k_*h)\over\sinh(2k_*h)+2k_*h} = {4\over 1 + {2k_*h\over\sinh(2k_*h)}},
\end{equation}
while
\begin{equation}
s(k_*) = {2\over gk_*(\tanh(kh) + k_*h \, \sech^2(kh))}.
\end{equation}
Then in the asymptotic regimes we have
\begin{equation}
d_{S,\mathrm{space}}(s\to \infty) = d_S(k_*\to 0) \approx 2; 
\quad
d_{S,\mathrm{space}}(s\to 0) = d_S(k_*\to \infty) \approx 4.
\end{equation}
So in the infrared one recovers the topological dimension, as expected --- but in the ultraviolet something slightly unusual happens in that the spectral dimension \emph{increases}. This is ultimately due to the fact that for large $k$ we have $\Omega(k)\propto k^{1/2}$, with an exponent that is lower than the $\Omega(k)\propto k$ behaviour characteristic of the infrared. 

Equivalently we could work with fixed $k_*$ and consider the effect of varying $h$. We have
\begin{equation}
d_{S,\mathrm{space}}(h\to 0) \approx 2; 
\quad
d_{S,\mathrm{space}}(h\to \infty) \approx 4.
\end{equation}
This implies that for shallow water surface waves  the spectral dimension is 2, whereas for deep water surface waves the spectral dimension is 4.

If you include surface tension the dispersion relation is modified and eventually at very short distances a different asymptotic behaviour dominates:  $\Omega \propto k^{3/2} $ at very large $k$. So in this region  the surface tension dominated deep water waves exhibit a spectral dimension:
\begin{equation}
d_{S,\mathrm{space}}(s\to0) \approx {d\over z} = {2\over 3/2} = {4\over3}. 
\end{equation}

\section{Conclusions}

We have analyzed the concept of the spectral dimension in the continuum and focused mainly on scales where curvature effects can be neglected. We have argued that in this regime the spectral dimension characterizes a differential operator that is used to define the (fictitious) diffusion process, or better yet the dispersion relation associated with this operator. We exhibited a natural way of assigning a spectral dimension to any dispersion relation, and shown how one can in principle invert the relationship. These results establish that the spectral dimension acts as probe of the kinematics of a given degree of freedom in this regime. 

We have considered some simple examples of dispersion relations for which one can analytically determine the spectral dimension. However, for more general (and realistic) dispersion relations this analytic determination is rarely feasible. To address this issue we have developed approximate techniques that allow one to determine the behaviour of the spectral dimension to a desired accuracy in various regimes of interests. In particular, we have presented asymptotics expansions (that resemble Seeley--DeWitt and Feynman expansions) which allow one to obtain the infrared and ultraviolet behavior of the spectral dimension (in flat space), and a technique to calculate the flow of the spectral dimension at intermediate scales using a saddle point approximation.

 As a major application of the theoretical and technical developments presented here, we consider the use of the concept of the spectral dimension as a link between discrete and continuum quantum gravity theories. Given that in the former the spectral dimension is one of the few known observables, defining it and providing the techniques for calculating in the latter is an important step in understanding the continuum limit of discrete models. In a companion article~\cite{CDT-Horava} we shall focus on this issue in more detail.
 
\addcontentsline{toc}{section}{Acknowledgments}
\section*{Acknowledgements}
TPS and SW were supported by Marie Curie Fellowships.   MV was supported by the Marsden Fund, administered by the Royal Society of New Zealand. 
We further acknowledge partial support via a FQXi travel grant. 

\addcontentsline{toc}{section}{References}



\begin{thebibliography}{99}

 \bibitem{CDT5}
  J.~Ambjorn, J.~Jurkiewicz, R.~Loll,
  ``Spectral dimension of the universe'',
  Phys.\ Rev.\ Lett.\  {\bf 95 } (2005)  171301.
  [hep-th/0505113].
  
 \bibitem{CDT1}
 J.~Ambjorn, A.~Gorlich, J.~Jurkiewicz, and R.~Loll,
  ``CDT --- an Entropic Theory of Quantum Gravity'',
  [arXiv:1007.2560 [hep-th]].
  
   \bibitem{CDT2}
J.~Ambjorn, J.~Jurkiewicz, R.~Loll,
  ``Causal Dynamical Triangulations and the Quest for Quantum Gravity'',
   [arXiv:1004.0352 [hep-th]].
   
  \bibitem{CDT3}
  J.~Ambjorn, A.~Gorlich, S.~Jordan, J.~Jurkiewicz, and R.~Loll, 
  ``CDT meets Horava-Lifshitz gravity'',
  Phys.\ Lett.\  {\bf B690 } (2010)  413-419.
  [arXiv:1002.3298 [hep-th]].
  
  \bibitem{CDT4}
 D.~Benedetti, R.~Loll, F.~Zamponi,
  ``(2+1)-dimensional quantum gravity as the continuum limit of Causal Dynamical Triangulations'',
  Phys.\ Rev.\  {\bf D76} (2007)  104022.
  [arXiv:0704.3214 [hep-th]].
  
  
  \bibitem{CDT6}
  D.~Benedetti, J.~Henson,
  ``Spectral geometry as a probe of quantum spacetime'',
  Phys.\ Rev.\  {\bf D80} (2009)  124036.
  [arXiv:0911.0401 [hep-th]].
 

  
   \bibitem{Benedetti:2008gu}
  D.~Benedetti,
  ``Fractal properties of quantum spacetime'',
  Phys.\ Rev.\ Lett.\  {\bf 102}, 111303 (2009)
  [arXiv:0811.1396 [hep-th]].

  
 \bibitem{Laiho:2011ya}
  J.~Laiho and D.~Coumbe,
  ``Evidence for Asymptotic Safety from Lattice Quantum Gravity'',
  arXiv:1104.5505 [hep-lat].
  
  
\bibitem{Atkin:2011ak}
  M.~R.~Atkin, G.~Giasemidis and J.~F.~Wheater,
  ``Continuum Random Combs and Scale Dependent Spectral Dimension'',
  J.\ Phys.\ A  {\bf 44}, 265001 (2011)
  [arXiv:1101.4174 [hep-th]].

 
   \bibitem{Horava}
P.~\Horava,
  ``Spectral Dimension of the Universe in Quantum Gravity at a Lifshitz Point'',
  Phys.\ Rev.\ Lett.\  {\bf 102 } (2009)  161301.
  [arXiv:0902.3657 [hep-th]].
 
 \bibitem{Carlip1}
 S.~Carlip,
  ``The Small Scale Structure of Spacetime'',
  arXiv:1009.1136 [gr-qc].
  
  
  \bibitem{Carlip2}
 S.~Carlip,
  ``Spontaneous Dimensional Reduction in Short-Distance Quantum Gravity?'',
  arXiv:0909.3329 [gr-qc].
  
  
    
\bibitem{Modesto:2009qc}
  L.~Modesto and P.~Nicolini,
  ``Spectral dimension of a quantum universe'',
  Phys.\ Rev.\ {\bf D81}, 104040 (2010)
  [arXiv:0912.0220 [hep-th]].
  
 
\bibitem{Pinzul:2010ct}
  A.~Pinzul,
  ``On spectral geometry approach to Horava-Lifshitz gravity: Spectral
  dimension'',
  arXiv:1010.5831 [hep-th].
  

 
 
 \bibitem{CDT-Horava}
T. P. Sotiriou, M. Visser, and   S. Weinfurtner,  
 ``Spectral dimensions as a probe of the ultraviolet continuum regime of causal dynamical triangulations". 
 Physical Review Letters {\bf107} (2011) 131303
[arXiv:1105.5646 [gr-qc]]

 
  \bibitem{Post}
  E.~Post, 
  ``Generalized differentiation'', 
  Transactions of the American Mathematical Society {\bf32} (1930) 723--781.
  
  \bibitem{Bryan}
  K.~Bryan, 
  ``Elementary inversion of the Laplace transform'',  2006. 
   {\sf http://www.rose-hulman.edu/~bryan/invlap.pdf}
  
    
 \bibitem{Horava1}
 P.~\Horava,
  ``Quantum Gravity at a Lifshitz Point'',
  Phys.\ Rev.\  {\bf D79 } (2009)  084008.
  [arXiv:0901.3775 [hep-th]].
 
  
  \bibitem{SVW1}
 T.~P.~Sotiriou, M.~Visser, S.~Weinfurtner,
  ``Quantum gravity without Lorentz invariance'',
  JHEP {\bf 0910 } (2009)  033.
  [arXiv:0905.2798 [hep-th]].
 
 \bibitem{SVW2}
 T.~P.~Sotiriou, M.~Visser, S.~Weinfurtner,
  ``Phenomenologically viable Lorentz-violating quantum gravity'',
  Phys.\ Rev.\ Lett.\  {\bf 102 } (2009)  251601.
  [arXiv:0904.4464 [hep-th]].
  
\bibitem{SVW3}
S.~Weinfurtner, T.~P.~Sotiriou, M.~Visser,
  ``Projectable \Horava--Lifshitz gravity in a nutshell'',
  J.\ Phys.\ Conf.\ Ser.\  {\bf 222 } (2010)  012054.
  [arXiv:1002.0308 [gr-qc]].

  
 \bibitem{LSB}
 M.~Visser,
  ``Lorentz symmetry breaking as a quantum field theory regulator'',
  Phys.\ Rev.\  {\bf D80 } (2009)  025011.
  [arXiv:0902.0590 [hep-th]].
  
  \bibitem{LSB2}
  M.~Visser,
  ``Power-counting renormalizability of generalized Horava gravity'',
  arXiv:0912.4757 [hep-th].
 
 
   
\bibitem{Blas}
D.~Blas, O.~Pujolas, S.~Sibiryakov,
  ``Consistent Extension of \Horava{} Gravity'',
  Phys.\ Rev.\ Lett.\  {\bf 104}, 181302 (2010).
  [arXiv:0909.3525 [he


\bibitem{2+1-in-preparation}  
T.~P.~Sotiriou, M.~Visser, S.~Weinfurtner, 
``Lower dimensional \Horava--Lifshitz gravity'', 
Phys. Rev. D {\bf83} (2011) 124021.
[arXiv:1103.3013 [hep-th]].
  
\bibitem{Sotiriou:2010wn}
  T.~P.~Sotiriou,
  ``Horava-Lifshitz gravity: a status report'',
  J.\ Phys.\ Conf.\ Ser.\  {\bf 283}, 012034 (2011)
  [arXiv:1010.3218 [hep-th]].

   
 \bibitem{compact}
 C.~Barcel\'o, S.~Liberati, S.~Sonego, and M.~Visser,
  ``Hawking-like radiation from evolving black holes and compact horizonless objects'',
  JHEP  {\bf1102 }(2011) 003 
  [arXiv:1011.5911 [gr-qc]].


\end{thebibliography}
\end{document}